\documentstyle[preprint,aps,epsf,tighten]{revtex}

\newcommand{\slashed}[1]{\rlap{$#1$}/}

\newcommand{\GeV}{\mbox{\rm GeV}}

\newcommand{\tr}{\mbox{\rm tr}}
\newcommand{\lsim}[1]{
\setlength{\unitlength}{12pt}
\begin{picture}(1.4,1.)
\put(.7,-0.3){\makebox(0.0,1.)[t]{$<$}}
\put(.7,-0.3){\makebox(0.0,1.)[b]{$\sim$}}
\end{picture}#1}

\begin{document}

\preprint{ZTF-99/02}

\title{
Schwinger-Dyson approach and generalized impulse approximation 
for the $\pi^0\gamma^\star\gamma$ transition
}

\author{Dubravko Klabu\v{c}ar}
\address{\footnotesize Department of Physics, Faculty of Science, \\
        Zagreb University, P.O.B. 162, 10001 Zagreb, Croatia}

\author{Dalibor Kekez}
\address{\footnotesize Rudjer Bo\v{s}kovi\'{c} Institute,
         P.O.B. 1016, 10001 Zagreb, Croatia}

\maketitle

\begin{abstract}
\noindent 
We review the pion-photon transition form factor calculated in the 
Schwinger-Dyson approach and an impulse approximation. We 
present results for it far above the scales presently accessible for
measurement, up to 36 GeV$^2$, and demonstrate agreement with the 
analytically inferred asymptotic behavior, for which we also provide 
a new derivation. We discuss how measurements at Jefferson Lab can 
provide information on how quarks are dynamically dressed. 
\end{abstract}

\vskip 1.5cm

\section{Introduction and survey}

A modern version of the constituent quark model with many attractive
features is one of remarkable achievements of the Schwinger-Dyson (SD) 
approach (reviewed in Refs.~\cite{NucTh9807026,RW}) to the physics of 
quarks and hadrons. 
It is well-known that most quark--antiquark bound state approaches have 
grave problems when they are faced with the electromagnetic processes 
dominated by Abelian axial anomaly. 
(See Ref. \cite{KeBiKl98} for a brief review and more references.)
However, these problems are resolved
in the SD approach thanks to its good chiral properties.
In the chiral limit, the axial-anomaly result for 
$\pi^0(p)\to\gamma(k)\gamma(k^\prime)$ transition amplitude
\begin{equation}
 T_{\pi^0}(0,0) =\frac{1}{4\pi^2 f_\pi} \,
\label{AnomAmpl}
\end{equation}
is reproduced analytically and exactly in this approach \cite{bando94,Roberts}.
The amplitude in Eq. (\ref{AnomAmpl}) is the limit when
both photons are real ($k^2 = {k^\prime}^2=0$), of the general
amplitude $T_{\pi^0}(k^2,{k^\prime}^2)$. 

At first, $\pi^0\to\gamma\gamma^{(\star)}$ processes and other applications 
were studied in the approach with
physically motivated {\it Ans\" atze} ({\it e.g.} \cite{Roberts,Frank+al}),
instead of the solutions of SD equations (SDE) with specified dynamics. 
However, the exact evaluation of Abelian axial anomaly in the closed
form (\ref{AnomAmpl}) occurs also in the variant where one actually solves 
SDE for the quark propagators and, in a consistent approximation and 
with the same model interaction, Bethe-Salpeter (BS) equations (BSE) for 
quark-antiquark bound states. This is the consistently coupled SD-BS 
approach, of which \cite{jain93b} and references therein provide an 
elaborated example. 

Since the anomaly is independent of hadronic structure, it is
essential to note that the relation (\ref{AnomAmpl}) is successfully 
reproduced in the SD approach irrespective of what concrete
{\it Ansatz} for the quark propagator 
        \begin{equation}
        S(q)
        =
        [ A(q^2)\slashed{q} - B(q^2) ]^{-1}~,
        \label{quark_propagator}
        \end{equation}
is used, or what concrete consistently coupled SD-BS solutions are 
employed for the dressed quark propagator (\ref{quark_propagator}) 
and the pion $q\bar q$ bound-state BS vertex $\Gamma_{\pi^0}$.

The successful treatment of the Abelian axial anomaly is possible in the 
SD approach because this approach incorporates the dynamical chiral 
symmetry breaking (D$\chi$SB) into the bound states consistently, so that 
the pion, although constructed as a $q\bar q$ bound state, appears as a
Goldstone boson in the chiral limit. 
With that, the bound-state descriptions of mesons are reconciled
with chiral requirements stemming both from QCD as underlying 
fundamental theory, and from phenomenology. The scenario is essentially
of the Nambu--Jona-Lasinio (NJL) type, but {\it without} its low cutoff.
Consequently, the quark-antiquark bound state mesons can finally provide 
adequate descriptions of the processes where the axial anomaly is important
({\it e.g.}, see \cite{bando94,Roberts,Frank+al,AR96,KeKl1,KlKe2,KeBiKl98}),
such as the two-photon processes of light pseudoscalars, of
which the $\pi^0\to\gamma\gamma$ decay is the cleanest
example of an anomalous electromagnetic decay.

Having gotten in hands the fully satisfactory result (\ref{AnomAmpl}) 
for the $\pi^0\to\gamma\gamma$, the next thing to explore was  
its one-off-shell-photon extension $T_{\pi^0}(k^2,0)$ in the SD approach. 
Frank {\it et al.} \cite{Frank+al} studied it using {\it Ans\" atze}
for dressed quark propagators (\ref{quark_propagator}), and then 
Ref.~\cite{KeKl1} did it in the consistently coupled SD-BS approach, 
utilizing the solutions of Ref.~\cite{jain93b}. However, both papers
called for additional studies in two respects. First, both
had employed the soft and chiral limit for the pion; {\it i.e.}, 
the BS vertex was approximated by its leading, ${\cal O}(p^0)$ piece: 
\begin{equation}
\Gamma_{\pi^0}(q,p) \approx \Gamma_{\pi^0}(q,0) = 
\gamma_5 \, \lambda^3 \, \frac{B(q^2)}{f_\pi}   \,\, .
\label{softLimBS}
\end{equation}
($\lambda^3$ is the third Gell-Mann matrix of flavor $SU(3)$.)
In addition, Refs.~\cite{Frank+al,KeKl1} could compare their 
results just with then only available data by CELLO \cite{behrend91}, 
at $Q^2 \lsim 2.5$ GeV$^2$.

Lately, however, the interest in the form factor 
$T_{\pi^0}(-Q^2,0)$ for the transition
$\gamma^\star(k)\gamma(k^{\prime})\to\pi^0(p)$ (where
$k^{2} = - Q^2 \neq 0$ is the momentum-squared of the
spacelike off-shell photon $\gamma^\star$), 
has again been growing for both experimental
(the new CLEO data \cite{gronberg98} and
plans for new TJNAF measurements \cite{gagasCEBAF}) and 
theoretical reasons -- {\it e.g.}, see \cite{H+Kinoshita98}.

At large values of $Q^2$, the $\gamma^\star\gamma\to\pi^0$
transition form factor should be adequately given by 
perturbative QCD (pQCD) - {\it e.g.} see 
\cite{BrodskyLepage,JakobKrollRaulfs96+Cao+al96,M+Rady97}.
Nevertheless, it is still not quite certain which $Q^2$
is sufficiently large. 
For example, according to Refs. \cite{Radyushkin+Rusk3},
just pQCD may still be not quite sufficient even at the highest of 
the presently accessible momenta, $Q^2 \lsim 10$ GeV$^2$.
The pQCD approaches start having problems as $Q^2$ decreases 
more and more into the nonperturbative domain. 
Also, see Sec. IV of the recent Ref. \cite{M+Rady97}
for clarifications how such 
approaches \cite{JakobKrollRaulfs96+Cao+al96} fail to reproduce 
the anomaly-induced value (\ref{AnomAmpl}) at $Q^2=0$.

On the other hand, the treatment of the axial anomaly is a 
strong point of the SD approach, as pointed out above.
What must then be explored in the coupled SD-BS approach is if the
large $Q^2$ behavior is satisfactory.
In particular, the comparison with the new CLEO data \cite{gronberg98}
-- at $Q^2$ up to 8 GeV$^2$ -- must be made.
Whether the large $Q^2$ behavior is satisfactorily close to the 
data and to the predictions of pQCD is a tricky question for a
constituent quark model. Namely, the calculation of the
transition form factor carried out in the simple constituent quark model
(with the constant light--quark mass parameter $m_u$), leads to
$T_{\pi^0}(-Q^2,0) \propto (m_u^2/Q^2)\ln^2(Q^2/m_u^2)$
as $Q^2\to\infty$, which overshoots both the CLEO data 
and pQCD predictions considerably \cite{H+Kinoshita98}. 
This is because of the additional $\ln^2(Q^2)$-dependence
on top of the large $Q^2$ behavior 
\begin{equation}
 T_{\pi^0}(-Q^2,0) = {\cal J} \, \frac{f_\pi}{Q^2} \,
\qquad ({\cal J} \to {\rm const} \,\, {\rm as \,\,} Q^2 \to \infty),
\label{largeQ2}
\end{equation}
favored by pQCD and other QCD-based theoretical predictions such as 
\cite{BrodskyLepage,JakobKrollRaulfs96+Cao+al96,M+Rady97,Radyushkin+Rusk3,ChernyakZhit,manohar90} and references therein.
Publication of the CLEO data \cite{gronberg98} made it clear that the
large-$Q^2$ behavior (\ref{largeQ2}) is also favored experimentally 
\cite{gronberg98}.

Fortunately, the constituent quark model provided by the SD approach, 
in which $B(q^2)/A(q^2)$ plays the role of the dynamically generated 
$q^2$-dependent mass, has turned out not to suffer from that shortcoming. 
Namely, we have shown in Ref.~\cite{KeKl3} that the above asymptotic 
behavior (\ref{largeQ2}) is in this approach obtained in the 
model-independent way.  

In this paper, we provide yet another way to obtain the
asymptotic $Q^2 \to \infty$ behavior found in Ref.~\cite{KeKl3}.
Predictions for $T_{\pi^0}(-Q^2,0)$ for finite 
values of $Q^2$ can be obtained by picking a definite model 
for interactions between quarks and obtaining corresponding
solutions of the SDE and BSE. In this paper, 
just as we did in Ref.~\cite{KeKl3}, we use the model of 
Ref.~\cite{jain93b}. However, here we give the
results over much wider range of $Q^2$ - see Fig. 1 
and the discussion thereof. Comparison with the data 
reveals in what way more precise measurements at intermediate
momenta, feasible at TJNAF, can both give insights in the
hadronic structure and provide guidance how to improve 
presently existing models in the SD-BS approach. 
In addition, the coupled SD-BS variant of the SD approach
(such as that in Ref.~\cite{jain93b}) can readily be extended
and applied beyond the soft and chiral limits [given by Eq.
(\ref{softLimBS})] for the bound-state vertex  - {\it e.g.}
see Refs.~\cite{KeKl1,MarisRoberts97PRC56,KlKe2,KeBiKl98}.
We did this when calculating $T_{\pi^0}(-Q^2,0)$ in Ref.~\cite{KeKl3},
but here we locate and point out the main reason for the difference
between the full calculation of the transition form factor 
and the one using the the soft and chiral limit approximation
(\ref{softLimBS}) for the pion BS vertex.

\section{Dressed quarks and BS-vertices need 
                dressed $\lowercase{qq}\gamma$ vertices}

In the coupled SD-BS approach, the BSE for the pion bound-state 
$q\bar q$ vertex $\Gamma_{\pi^0}(q,p)$ employs the dressed quark 
propagator (\ref{quark_propagator})
obtained by solving its SDE. Solving the SDE and BSE 
in a consistent approximation is crucial ({\it e.g.}, see Refs. 
\cite{jain91,munczek92,jain93b,Bender+al96,MarisRoberts97PRC56})
for obtaining $q\bar q$ bound states which are, in the case of light 
pseudoscalar mesons, simultaneously also the (pseudo-)Goldstone bosons of 
D$\chi$SB.

Following Jain and Munczek \cite{jain91,munczek92,jain93b}, we 
adopt the ladder-type approximation
sometimes called the {\it improved} \cite{Miransky} or {\it generalized} 
\cite{Roberts} ladder approximation (employing bare quark--gluon--quark 
vertices but dressed propagators). For the gluon propagator we use an effective,
({\it partially}) modeled one in Landau-gauge \cite{jain91,munczek92,jain93b}, 
given by
$G(-l^2)( g^{\mu\nu} - {l^\mu l^\nu}/{l^2} )~.$
(This Ansatz is often called the 
``Abelian approximation" \cite{MarisRoberts97PRC56}.)
The effective propagator function $G$ is the
sum of the perturbative contribution $G_{UV}$ and the nonperturbative
contribution $G_{IR}$:
$G(Q^2) = G_{UV}(Q^2) + G_{IR}(Q^2)~,\;\;(Q^2 = -l^2)~.$
The perturbative part $G_{UV} = (16\pi/3)\alpha_s(Q^2)/Q^2$ 
is well-known from perturbative QCD, so it is {\it not} modeled
\cite{jain91,munczek92,jain93b,KeKl1,KlKe2,KeBiKl98}.
As in Refs. \cite{KeKl1,KlKe2,KeBiKl98}, we follow 
Refs. \cite{jain91,munczek92,jain93b} and employ the 
two--loop asymptotic expression for $\alpha_s(Q^2)$.
For the modeled, IR part of the gluon propagator, we 
adopt from Ref.~\cite{jain93b}
$G_{IR}(Q^2)=(16\pi^2/3) \,a\,Q^2 e^{-\mu Q^2}$, 
with their parameters
$a=(0.387\,\GeV)^{-4}$ and $\mu=(0.510\,\GeV)^{-2}$.
For details of how we solve SDE and BSE, we refer to Refs. 
\cite{KeKl1,KlKe2,KeBiKl98}. To high accuracy, we reproduce 
Jain and Munczek's \cite{jain93b} solutions of SDE for the dressed 
propagators $S(q)$, i.e., the functions $A(q^2)$ and $B(q^2)$, 
as well as the BS solutions for the four functions comprising 
the pion bound-state vertex $\Gamma_{\pi^0}$. 

The $\pi^0\gamma^\star\gamma$ transition tensor ($T_{\pi^0}^{\mu\nu}$)
and scalar ($T_{\pi^0}$) amplitudes, related by
        \begin{equation}
        T_{\pi^0}^{\mu\nu}(k,k^\prime)
        =
        \varepsilon^{\alpha\beta\mu\nu} k_\alpha k^\prime_\beta
        T_{\pi^0}(k^2,k^{\prime 2})~,
        \label{Tmunu-parametrization}
        \end{equation}
are found in the present paper as in Refs. \cite{KeKl1,KlKe2,KeBiKl98}.
That is, the pseudoscalar-vector-vector (PVV) triangle graph is 
calculated by employing the framework advocated by (for example)
\cite{bando94,Roberts,Frank+al,Burden+al,AR96}
in the context of electromagnetic interactions of BS bound states,
and often called the generalized impulse approximation (GIA) -
{\it e.g.}, by \cite{Frank+al,Burden+al}.
To evaluate the triangle graph, we therefore use the {\it dressed} 
quark propagator $S(q)$ and the pseudoscalar 
($P=\pi^0, \eta_8, \eta_0, \eta_c ...$) meson BS bound--state 
vertex $\Gamma_P(q,p)$. (See Fig.~1 in Ref.~\cite{KeKl3}, for example.)
Equivalently, we can work with the BS-amplitude 
$\chi_{P}(q,p)\equiv S(q+{p}/{2})\Gamma_{P}(q,p)S(q-{p}/{2})$.
This is in fact what we do, following Ref. \cite{jain93b}. 

The third ingredient crucial for GIA's ability to reproduce the 
correct Abelian anomaly result (\ref{AnomAmpl}),
is employing in the triangle graph appropriately dressed
{\it electromagnetic} vector vertices $\Gamma^\mu(q^\prime,q)$,
which satisfy the vector Ward--Takahashi identity (WTI)
$(q^\prime-q)_\mu \Gamma^\mu(q^\prime,q)=S^{-1}(q^\prime)-S^{-1}(q)~.$
Namely, assuming that photons
couple to quarks through the bare vertex $\gamma^\mu$
would be inconsistent with
our quark propagator $S(q)$, which, dynamically dressed through
its SD-equation, contains the momentum-dependent
functions $A(q^2)$ and $B(q^2)$.
The bare vertex $\gamma^\mu$ obviously violates the vector WTI, implying 
the non-conservation of the electromagnetic vector current and 
charge.
Solving the pertinent SD equation for the dressed quark--photon--quark 
($qq\gamma$) vertex $\Gamma^\mu$ is still an unsolved problem, and using 
the realistic Ans\"{a}tze for $\Gamma^\mu$ is presently the only practical 
way to satisfy the WTI. 
Following Refs. \cite{Frank+al,Burden+al,Roberts,AR96} (for example), 
we can choose the Ball--Chiu (BC) ~\cite{BC} vertex; {\it i.e.}, 
$\Gamma^\mu = \Gamma^\mu_{BC}$,

        \begin{eqnarray}
        \Gamma^\mu_{BC}(q^\prime,q) =
        A_{\bf +}(q^{\prime 2},q^2)
       \frac{\gamma^\mu}{\textstyle 2}
        + \frac{\textstyle (q^\prime+q)^\mu }
               {\textstyle (q^{\prime 2} - q^2) }
        \{A_{\bf -}(q^{\prime 2},q^2)
        \frac{\textstyle (\slashed{q}^\prime + \slashed{q}) }{\textstyle 2}
         - B_{\bf -}(q^{\prime 2},q^2) \}~,
        \label{BC-vertex}
        \end{eqnarray}
where
$H_{\bf \pm}(q^{\prime 2},q^2)\equiv [H(q^{\prime 2})\pm H(q^2)]$,
for $H = A$ or $B$.
This particular solution of the vector WTI 
reduces to the bare 
vertex in the free-field limit as must be in perturbation theory, has 
the same transformation properties under Lorentz transformations and 
charge conjugation as the bare vertex, and has no kinematic singularities. 
It does {\it not} introduce any new parameters as it is completely
determined by the dressed quark propagator $S(q)$. 

Another WTI-preserving choice for $\Gamma^\mu$ can be a vertex of 
the Curtis--Pennington (CP) type, {\it i.e.},
$\Gamma^\mu \equiv \Gamma^\mu_{BC} + \Delta \Gamma^\mu$
where the the transverse addition $\Delta \Gamma^\mu$ is of 
the type \cite{CP90} 
\begin{equation}
\Delta\Gamma^\mu(q^\prime,q) = 
\frac{\gamma^\mu(q^{\prime 2} - q^2)-(q^\prime + q)^\mu
                ( \slashed{q}^\prime - \slashed{q} )}
             {2d(q^\prime,q)} {A_{\bf -}(q^{\prime 2},q^2)} \, .
\label{defDelta}
\end{equation}
Two especially suitable {\it Ans\" atze} for the dynamical function
$d(q^\prime,q)$, ensuring multiplicative renormalizability of fermion SDE
beyond the ladder approximation in QED$_4$, are
\begin{equation}
d_\pm(q^\prime,q) = 
        \frac{1}{q^{\prime 2}+q^2}
        \left\{
                (q^{\prime 2} \pm q^2)^2
                +
                \left[
                        \frac{B^2(q^{\prime 2})}{A^2(q^{\prime 2})}
                        +
                        \frac{B^2(q^2)}{A^2(q^2)}
                \right]^2
        \right\}~.
\label{defdfunct}
\end{equation}
The original CP {\it Ansatz} $\Gamma^\mu \equiv \Gamma^\mu_{CP}$
employed $d(q^\prime,q)= d_-(q^\prime,q)$ \cite{CP90}.
We will use it in analytic calculations of
$T(-Q^2,0)$, which are possible for $Q^2=0$ and $Q^2\to \infty$.
However, in the numerical calculations, which are
necessary for finite values of $Q^2\neq 0$, we prefer to 
restrict ourselves (besides the minimal BC vertex) to the
{\it modified} CP (mCP) vertex, $\Gamma^\mu_{mCP}$, resulting from
the choice $d = d_+$ in Eq. (\ref{defDelta}).
Namely, as pointed out in Ref. \cite{KeKl3}, the numerical calculation
of $T(-Q^2,0)$ employing $\Gamma^\mu_{mCP}$ vertices
is free from certain numerical difficulties arising,
in this application, from the CP denominator function $d_-(q^\prime,q)$.

In contrast to the BC one, the mCP vertex is also consistent
with multiplicative renormalizability, like the original CP vertex.
In the present context,
the important {\it qualitative} difference between the BC-vertex on
one side, and the CP vertex as well as the modified, mCP vertex
on the other side, will be that
$\Gamma^\mu_{BC}(q^\prime,q)\to\gamma^\mu$
when {\it both} $q^{\prime 2},q^2\to \pm \infty$, whereas
$\Gamma^\mu_{CP}(q^\prime,q) \to \gamma^\mu$ and
$\Gamma^\mu_{mCP}(q^\prime,q) \to \gamma^\mu$
as soon as {\it one} of the squared momenta tends to infinity.
This turned out \cite{KeKl3} to lead to the same coefficient 
of the $Q^2 \to \infty$ behavior (\ref{largeQ2}) for the 
CP vertices and mCP vertices,
namely ${\cal J}=4/3$, but a larger one for the BC-vertices.

We have checked that the $Q^2 = 0$ case, {\it i.e.}, the 
$\pi^0\to\gamma\gamma$ amplitude (\ref{AnomAmpl}), is reproduced 
analytically employing the CP vertices and mCP vertices in the 
same way as when the BC vertices were employed in earlier 
applications, {\it e.g.} \cite{Roberts,Frank+al,KeKl1,KlKe2}.

In the case of $\pi^0$, GIA yields ({\em e.g.}, see Eq. (24) in 
Ref. \cite{KlKe2}) the amplitude $T_{\pi^0}^{\mu\nu}(k,k^\prime)$:
\begin{displaymath}
        T_{\pi^0}^{\mu\nu}(k,k^\prime)
        =
        -
        N_c \,
        \frac{1}{3\sqrt{2}}
        \int\frac{d^4q}{(2\pi)^4} \mbox{\rm tr} \{
        \Gamma^\mu(q-\frac{p}{2},k+q-\frac{p}{2})
        S(k+q-\frac{p}{2})
\end{displaymath}
\begin{equation}
          \qquad
        \times
        \Gamma^\nu(k+q-\frac{p}{2},q+\frac{p}{2})
        \chi(q,p) \}
        +
        (k\leftrightarrow k^\prime,\mu\leftrightarrow\nu).
\label{Tmunu(2)}
\end{equation}
Here, $\chi$ is the BS amplitude of both $u\bar u$ and $d\bar d$
pseudoscalar bound states: $\chi \equiv \chi_{u\bar u} = \chi_{d\bar d}$
thanks to the isospin symmetry assumed here. This symmetry likewise
enables us to continue suppressing flavor labels also on the quark 
propagators $S$ and $qq\gamma$ vertices $\Gamma^\mu$.
We follow the conventions of Ref. \cite{KlKe2}, including those 
for the flavor factors and flavor matrices $\lambda^a$. Then, 
$\chi_{\pi^0}(q,p) \equiv \chi(q,p) \, \lambda^3/\sqrt{2}$, so that 
the prefactor $1/3\sqrt{2}$ in Eq. (\ref{Tmunu(2)}) is just the flavor 
trace ${\rm tr}({\cal Q}^2\lambda^3/\sqrt{2})$ where 
${\cal Q}=\lambda^3/2 + \lambda^8/2\sqrt{3}=diag(+2/3, -1/3, -1/3)$ 
is the quark charge matrix.

\section{Our results and comparison with others}

Already in Ref. \cite{KeKl1} (where only the BC vertex was employed), the 
transition form factor $T_{\pi^0}(-Q^2,0)$ was numerically evaluated 
(for $0 < Q^2 < 2.8$ GeV$^2$ only) employing - for the pion only - the 
soft and chiral limit (\ref{softLimBS}).

In the subsequent work \cite{KeKl3}, we went  
beyond this approximation, using our complete solution for the BS vertex 
$\Gamma_{\pi^0}(q,p)$, {\it viz.} the BS amplitude $\chi_{\pi^0}(q,p)$, 
given by the decomposition into 4 scalar functions multiplying independent 
spinor structures. We calculated the transition form factor not only
for the BC vertex, but also for the mCP vertex. The approximation we kept in 
Ref.~\cite{KeKl3} (and which we still keep in the present paper), is 
discarding the second and higher derivatives in the momentum expansions
of the SD solutions $A(q^2)$ and $B(q^2)$ and BS solutions $\chi(q,p)$.  
In Ref.~\cite{KeKl3}, $T_{\pi^0}(-Q^2,0)$ was evaluated up to $Q^2
\approx 8$ GeV$^2$, which is roughly the limit of the presently 
accessible transferred momenta. In this paper, we evaluate the 
transition form factor up to 36 GeV$^2$. We present 
curves for $Q^2 T_{\pi^0}(-Q^2,0)$ evaluated in various ways first
in Fig. 1, depicting their evolution all the way to 36 GeV$^2$.
This squeezes the presently existing experimental points into the first 
quarter of the scale; however, the issue of comparing and improving 
the agreement with the data at the accessible momenta will be returned
to later, and now we want first to illustrate how various 
theoretical transition form factors at finite $Q^2$ connect to
the asymptotic behavior.

In Fig. 1, the solid line is our prediction for $Q^2 T_{\pi^0}(-Q^2,0)$
evaluated with the BC vertex. It reaches the asymptotic behavior faster
than other curves. Already after $Q^2 \approx 4$ GeV$^2$, the tiny
changes cannot be observed in the solid line, but only in the corresponding
numerical data, as the asymptotic value is approached from below.
For comparison, the dash-dotted line gives  
the results obtained (also with the BC vertex) in the approximation 
(\ref{softLimBS}) of soft and chiral limit as explained above. 
At $Q^2 \approx 4$ GeV$^2$, where the $Q^2 T_{\pi^0}(-Q^2,0)$ from the
full calculation already practically reaches the asymptotic limit,
the difference between it and $Q^2 T_{\pi^0}^{soft}(-Q^2,0)$ from
the approximation (\ref{softLimBS}), is somewhat above 10\%. This 
difference gets bigger as $Q^2$ grows, and asymptotically it reaches some 
22\% of the limit value of $Q^2 T_{\pi^0}(-Q^2,0)$. However, this still 
means that the effect of the soft limit approximation on the transition
form factor is much less than on the charge form factor. Namely,
Maris and Roberts \cite{NTh9804062} pointed out that omitting the 
pseudovector components of the pion lead even to the wrong asymptotic
behavior, as $1/Q^4$ instead of $1/Q^2$, in the case of the charge 
form factor of the pion. In the present case of the transition form 
factor, the correct $1/Q^2$ leading behavior is nevertheless obtained 
not only for the results of the full calculation, but also in the 
soft limit approximation (\ref{softLimBS}). It is just that the 
coefficient of $1/Q^2$ is underestimated with respect to the full 
calculation.

The dashed line was obtained in the same way as the solid line, but 
employing the mCP vertex. Even at $Q^2 = 36$ GeV$^2$, the asymptotic 
behavior is obviously not yet reached in the case of the mCP vertex. 
However, later we will comment on how Ref.~\cite{KeKl3} showed 
analytically that the behavior given by Eq. (\ref{largeQ2}) must be 
reached at some point, although at much higher values of $Q^2$ when 
the mCP vertex is used.

How do our transition form factors agree with other theoretical 
approaches? The vector--meson dominance (VMD) model and the QCD sum 
rule approach \cite{Radyushkin+Rusk3} give the transition form factor 
which is some 10\% below our ``BC" $T_{\pi^0}(-Q^2,0)$ for the 
{\it presently} largest accessible values of $Q^2$, {\it i.e.}, around 
8 GeV$^2$. Therefore, in that region, our BC-results are 
between the uppermost line in Fig. 1 (the dotted constant line at 
$2 f_\pi$) denoting the asymptotic pQCD \cite{BrodskyLepage} version 
(${\cal J} = 2$) of Eq. (\ref{largeQ2}), and the results of VMD 
({\it e.g.}, see \cite{H+Kinoshita98}), the recent pQCD calculation 
by Ref.~\cite{Brodsky+al98} as well as the QCD sum rule results
of Radyushkin and Ruskov \cite{Radyushkin+Rusk3}.
$Q^2 T_{\pi^0}(-Q^2,0)$
of VMD and pQCD rise with $Q^2$, albeit with different rates, while
that from the QCD sum rules of Ref.~\cite{Radyushkin+Rusk3} starts
almost imperceptibly falling after $Q^2 \approx 7$ GeV$^2$.

The large-$Q^2$ leading power-law behavior (\ref{largeQ2}) was 
first derived from the parton picture in the 
infinite momentum frame -- {\it e.g.}, see Ref. \cite{BrodskyLepage}.
In this and other similar pQCD approaches, 
the precise value of the coefficient of the leading $1/Q^2$ term
depends on the pion distribution amplitude 
$\varphi_\pi(x)$ which should contain the necessary nonperturbative
information about the probability that a partonic quark carries the
fraction $x$ of the total longitudinal momentum. 
{\it E.g.}, the well-known example of a ``broad" distribution
$\varphi^{CZ}_\pi(x)= f_\pi 5 \sqrt{3} x(1-x)(1-2x)^2$ 
(proposed by Chernyak and Zhitnitsky \cite{ChernyakZhit} motivated 
by sum-rule considerations) leads to ${\cal J}=10/3$, but it is too 
large in the light of the latest data \cite{gronberg98}. In 
contrast, the asymptotic distribution 
$\varphi^{A}_\pi(x)= f_\pi \sqrt{3} x(1-x)$ 
(depicted by dotted line in Fig.~2 and favored by Lepage and Brodsky 
\cite{BrodskyLepage}) yields ${\cal J}=2$, resulting in 
the constant dotted line crossing the upper error bar of the presently 
highest-$Q^2$ data point in Fig. 1. 

In the {\it strict} $\ln(Q^2)\to\infty$ limit, every 
distribution amplitude must evolve into the asymptotic one,
$\varphi_\pi(x) \to \varphi^{A}_\pi(x)$, if the effects of the 
pQCD evolution are taken into account \cite{BrodskyLepage}. 
However, even at $Q^2$-values larger than the presently
accessible ones, other effects may still be more important than the
effects of the pQCD evolution. This is the reason why other approaches
and other forms of $\varphi_\pi(x)$ should be considered even when they
do not incorporate the pQCD evolution.
This is in line with Radyushkin and Ruskov's (\cite{Radyushkin+Rusk3}
and their references therein) pointing out desirability of having
direct calculations of $T_{\pi^0}(-Q^2,0)$ without {\it a priori} 
assumptions about the pion distribution amplitude $\varphi_\pi$.
One can then consider the opposite procedure from the one which is 
standard in pQCD: from such direct calculations of 
$T_{\pi^0}(-Q^2,0)$, one can draw conclusions about the 
distribution amplitude $\varphi_\pi$.   

The form 
\begin{equation}
\varphi_\pi(x) = \frac{f_\pi}{2 \sqrt{3}} 
              \frac{\Gamma(2\zeta+2)}{[\Gamma(\zeta+1)]^2}
               \, x^\zeta \, (1-x)^\zeta \, , 
           \quad \zeta > 0 \, ,
\label{genForm}
\end{equation}
is suitable for representing various distribution amplitudes because 
it is relatively general \cite{BrodskyLepage}: 
$\zeta > 1$ yields the distributions that are 
``peaked" or ``narrowed" with respect to the asymptotic one ($\zeta = 1$), 
whereas $\zeta < 1$ gives the ``broadened" distributions which, however, 
now seems to be ruled out for the same reason as $\varphi^{CZ}_\pi(x)$ 
quoted above, since ${\cal J} > 2$ is ruled out empirically by CLEO 
\cite{gronberg98}. Namely, it is easy to see that Eq.~(\ref{genForm}) 
implies $\zeta = 2/(3 {\cal J} - 4)$. 

In Fig.~2, we plot the distribution amplitudes of the form
(\ref{genForm}) for the three cases that are the most interesting
for the present discussion. As seen below, these cases
correspond to the $\zeta$-values equal to 1 (dotted line), 
1.5 (solid line), and 2.5 (dash-dotted line).

At the highest presently accessible momenta, the asymptotic 
prediction (${\cal J} = 2$) is
lowered by some 20\% by the lowest order QCD radiative corrections
\cite{Brodsky+al98}, amounting to ${\cal J}\approx 1.6$, which fits the 
CLEO data well. Of course, these QCD corrections mean 
that $Q^2 T_{\pi^0}(-Q^2,0)$ is not strictly constant, but according to 
Ref. \cite{Brodsky+al98} it rises towards $2 f_\pi$ so slowly 
that we can take it constant in practice.
[Eqs. (19) and (18) from Ref.~\cite{Brodsky+al98} lead to 
the line of empty squares in Fig.~1, that is, to
$Q^2 T_{\pi^0}(-Q^2,0)$ which grows just 4\% from $Q^2=9$ GeV$^2$ 
to $Q^2=36$ GeV$^2$ when the number of relevant flavors is set to 
$N_f=5$. The full calculation \cite{Brodsky+al98}, which employs the 
momentum-dependent $N_f$ \cite{Shirkov+Mikhailov94}, seems to give
even slower variation of $Q^2 T_{\pi^0}$ -- see their Fig.~2.]
Similarly slow variation results from the sum-rule approach of Radyushkin 
and Ruskov \cite{Radyushkin+Rusk3}, which yields the transition form factor 
which is, on the interval from 3 to 8 GeV$^2$, quite close to those of the 
pQCD approach of Brodsky {\it et al.} \cite{Brodsky+al98}. 
The sum-rule \cite{Radyushkin+Rusk3} $Q^2 T_{\pi^0}(-Q^2,0)$ starts
actually {\it falling} after $Q^2\sim 7$ GeV$^2$, but so slowly that 
Eq. (\ref{largeQ2}) with the constant ${\cal J}\approx 1.6$ represents
it accurately at the presently accessible values of $Q^2$. 
This, just like roughly the same ${\cal J}$ associated to 
Ref. \cite{Brodsky+al98}, corresponds to a rather narrow 
distribution (\ref{genForm}) with $\zeta \approx 2.5$,
depicted by the dash-dotted line in Fig.~3. 

The leading large-$Q^2$ behavior as in Eq. (\ref{largeQ2}), was obtained 
also by Manohar \cite{manohar90} using the operator product expansion (OPE). 
According to his OPE calculation, the coefficient in Eq. (\ref{largeQ2}) 
giving the leading term is ${\cal J}=4/3$, which is below our large-$Q^2$ 
$T_{\pi^0}(-Q^2,0)$ by $\sim 20\%$ when we use the BC vertex, but  
exactly coincides with the $Q^2\to\infty$ limit obtained 
when we use the CP or the mCP vertex -- see Ref. \cite{KeKl3} 
and comments below. 
The coefficient ${\cal J}=4/3$ is the lowest one still consistent with the 
form (\ref{genForm}) because it corresponds to $\zeta = \infty$. The pion 
distribution amplitude (\ref{genForm}) then becomes infinitely peaked 
delta function: $\varphi_\pi(x)=({f_\pi}/{2 \sqrt{3}})\, \delta(x - 1/2)$.

For $Q^2 > 4$ GeV$^2$, our ``BC" $T_{\pi^0}(-Q^2,0)$ also 
behaves in excellent approximation as (\ref{largeQ2}), with 
${\cal J} \approx 1.78$. This would in the pQCD factorization approach 
correspond to $\varphi_\pi(x)$ (\ref{genForm}) with $\zeta \approx 1.5$. 
On the other hand, our ``mCP" $T_{\pi^0}(-Q^2,0)$ falls off
faster than $1/Q^2$ for even the largest of the $Q^2$ values 
depicted in Fig. 2. However, it does not fall off much faster,
as our ``mCP" $Q^2 T_{\pi^0}(-Q^2,0)$ at $Q^2=18$ GeV$^2$ is only 6\%, 
and at the huge $Q^2=36$ GeV$^2$ is only 10\% smaller than at $Q^2=9$ 
GeV$^2$ (roughly the highest presently accessible $Q^2$).
Moreover, Ref.~\cite{KeKl3} showed analytically that, generally,  
the $1/Q^2$-behavior of Eq. (\ref{largeQ2}) is at some point 
reached in our approach, although Fig. 1 shows that for the mCP 
$qq\gamma$ vertices this can happen only at significantly higher 
$Q^2$ than it happens for the BC $qq\gamma$ vertices.

\section{ Discussion, comments and conclusions } 

The most important novelty in Ref.~\cite{KeKl3} is its result on the 
large-$Q^2$ asymptotic behavior. Namely, it implies that the modern 
version of the constituent quark model which is given by 
the coupled SD-BS approach, provides from $Q^2=0$ to $Q^2\to \infty$ the
description for $\gamma\gamma^\star\to\pi^0$ which is --
independently of model details -- consistent not only with the Abelian axial
anomaly but also with the QCD predictions (\ref{largeQ2}) for the leading
large-$Q^2$ behavior. In the SD-BS approach, $f_\pi$ is a quantity 
calculable through the straightforward application of the 
Mandelstam-formalism expression
\begin{equation}
p_\mu f_\pi  = i \frac{N_c}{\sqrt{2}}
\int\frac{d^4q}{(2\pi)^4}
\tr [\chi(q,p) \gamma_\mu \gamma_5] \, .
\label{fpi}
\end{equation}
The model dependence is present in the (successfully reproduced
\cite{jain93b,KeKl1,KlKe2,KeBiKl98}) {\it value} of $f_\pi$, but
the derivation \cite{KeKl3} of the asymptotic {\it forms} (\ref{largeQ2}) 
and (\ref{ampWtilde}) below, is model-independent. 
They do not depend on what are the bound-state solutions, just like
the chiral-limit axial-anomaly amplitude (\ref{AnomAmpl}) doesn't. 

Of special importance is also that the derivation in Ref.~\cite{KeKl3}
of the asymptotic large-$Q^2$ forms (\ref{largeQ2}) and (\ref{ampWtilde}) 
below, applies to both Minkowski and Euclidean space.

Our result \cite{KeKl3} on the asymptotic behavior of
$T_{\pi^0}(-Q^2,0)$ subsequently received further support
from Roberts \cite{RobertsDubr}, who generalized its derivation 
by taking into account renormalization explicitly. His derivation 
shows that the asymptotics of Ref.~\cite{KeKl3} obtained using the 
CP or mCP vertex, namely ${\cal J}=4/3$, must be precisely reproduced 
for any $qq\gamma$ vertex which is (like the CP or mCP ones) 
consistent with multiplicative renormalizability.

It is interesting that the asymptotic behavior for large negative 
$k^{2}=-Q^2$ predicted \cite{KeKl3} for  
the bare $qq\gamma$ vertex and the dressed CP and mCP ones, 
is in exact agreement with the leading term predicted 
by Manohar using OPE \cite{manohar90}. In this paper,
we shed some light on this connection by providing the
following alternative derivation of the asymptotic behavior:
consider $TJ^\mu_a(x)J^\nu_b(x)$, the $T$-product of two quark 
vector currents 
$J^\mu_a(x)={\bar \psi}(x)\gamma^\mu (\lambda^a/2) \psi(x)$, 
along the lines of Ref. \cite{Yndurain} (Ch. 
18 on OPE), but evaluated between the pion state and the vacuum.
We substitute for the quark propagator its leading light cone
[$(x-y)^2\to 0$] behavior
\begin{equation}
S(x-y) \approx 
\frac{2 (x-y)\cdot\gamma}{(2\pi)^2 [(x-y)^2 -i\epsilon]^2} \, ,
\label{LightConeS}
\end{equation}
and utilize 
$\gamma^\mu\gamma^\lambda\gamma^\nu =
{\cal S}^{\mu\lambda\nu\sigma} \gamma_\sigma
- i \varepsilon^{\mu\lambda\nu\sigma} \gamma_\sigma \gamma_5$
similarly as in the derivation of the large $Q^2$-behavior in 
Ref.~\cite{KeKl3}.
The term which contains the symmetric tensor
${\cal S}^{\mu\lambda\nu\sigma} = g^{\mu\lambda} g^{\nu\sigma}
 + g^{\mu\sigma} g^{\lambda\nu} + g^{\mu\nu} g^{\lambda\sigma}$,
is readily seen not to contribute to the $\pi^0\gamma\gamma^\star$ 
vertex. On the other hand, in the term containing 
the antisymmetric Levi-Civita tensor $\varepsilon^{\mu\lambda\nu\sigma}$
and $\gamma_5$, one finds -- in the lowest order of the $(x-y)$-expansion 
-- the pion-to-vacuum matrix element of the axial current,
\begin{equation}
\langle 0 |{\bar\psi}(0)\gamma_\mu\gamma_5{\cal Q}^2\psi(0)|\pi^3(p)\rangle
= \frac{1}{3} \langle 0|{\bar\psi}(0)\gamma_\mu\gamma_5
\frac{\lambda^3}{2}\psi(0)|\pi^3(p)\rangle 
= \frac{1}{3} \, i \, f_\pi \, p_\mu    \, ,
\end{equation}
defining the pion decay constant $f_\pi$. 
The coordinate-space integration is reduced to 
\begin{equation}
\int d^4x \, e^{ik\cdot x} \frac{x_\lambda}{(x^2 - i\epsilon)^2} 
= 2\pi^2 \,  \frac{k_\lambda}{k^2 + i\epsilon} \, .
\end{equation}
Putting all this together reveals that 
the lowest-order result in the $(x-y)$-expansion is just 
Eq. (\ref{largeQ2}) with ${\cal J}$ being precisely 4/3.

We thus in another way recover the result of Manohar \cite{manohar90},
of Ref.~\cite{KeKl3} for the case when the bare ($\gamma_\mu$) or 
the dressed CP or mCP vertices are used, and of subsequent generalization 
\cite{RobertsDubr} thereof for any $qq\gamma$ vertices consistent 
with the multiplicative renormalization.
However, this asymptotic behavior $Q^2 T_{\pi^0}(-Q^2,0) = (4/3) f_\pi $
lies some 20\% below the central values of the largest-$Q^2$
CLEO data, and is more than one standard deviation away from them.
Nevertheless, Manohar \cite{manohar90} pointed out that his
OPE approach also indicates the existence of potentially large 
corrections to his leading term. 

Let us now return to the BC vertex, which has been the most commonly 
used dressed $qq\gamma$ vertex in the phenomenological SD-BS calculations
so far. In conjunction with our SD-BS solutions, the usage 
of the BC vertex raises the asymptotic coefficient ${\cal J}$ of 
Eq.~(\ref{largeQ2}) from 4/3 resulting from the usage
of the bare, CP and mCP vertices, to ${\cal J}\approx 1.78$.
The reason for this enhancement is 
that the BC vertex does not reduce to the bare vertex $\gamma^\mu$
if the large momenta flow through just one, and not both, of its 
fermion legs. (Also note that the BC vertex is not consistent with 
multiplicative renormalizability \cite{BrownDorey91,CP90}, so that 
the arguments of Ref. \cite{RobertsDubr} are not applicable to it.) 
It was shown in Ref.~\cite{KeKl3} that when the BC vertex is used
in our approach, the predicted asymptotic behavior is
\begin{equation}
T_{\pi^0}(-Q^2,0) \to \frac{4}{3} \frac{{\widetilde f}_\pi}{Q^2}
\label{ampWtilde}
\end{equation}
where ${\widetilde f}_\pi$ is given by the same expression (\ref{fpi})
as $f_\pi$, except that the integrand is modified by the factor 
$[1+A(q^2)]^2/4$. In the special case of the solutions we use, the slow 
and moderate variation of $A(q^2)$, as well as its shape (see Fig. 3 in 
Ref. \cite{KeBiKl98}) permits the approximate factorization 
\begin{equation}
{\widetilde f}_\pi \approx f_\pi \frac{[1+A(0)]^2}{4}.
\label{approxFactor}
\end{equation}

This last approximation (a rather rough one) would imply 
${\cal J} \approx 1.69$, {\it i.e.}, $Q^2 T_{\pi^0}(-Q^2,0) \to 0.157$ GeV.
The more accurate Eq. (\ref{ampWtilde}) gives 
$Q^2 T_{\pi^0}(-Q^2,0) \to 0.168$ GeV. 
This is practically indistinguishable from the asymptotics indicated
by our full numerical calculation of the transition form factor, 
which reached $Q^2 T_{\pi^0}(-Q^2,0) \approx 0.165$ GeV at the largest
numerically studied $Q^2$. 
It is of special importance that analytically obtained large-$Q^2$
asymptotics (\ref{ampWtilde}), the derivation of which is valid in 
both Minkowski and Euclidean space, makes this smooth and accurate 
contact between our large-$Q^2$ numerical results calculated with the 
Euclidean solutions of the chosen model \cite{jain93b}. The excellent 
agreement between the analytical and numerical results confirms the 
accuracy of our numerical methods and procedures (employing the Euclidean 
bound-state solutions \cite{jain93b}) used in this and earlier 
papers~\cite{KeKl1,KlKe2,KeBiKl98}.

Another case of this agreement between the numerical results and the 
analytical considerations is provided by $Q^2 T_{\pi^0}(-Q^2,0)$ found
in the soft and chiral limit approximation (\ref{softLimBS}). Therefore,
let us for a moment consider again this approximation, although we 
surpassed it already in Ref.~\cite{KeKl3}. (As a bonus, we also gain
understanding of some subtle aspects in its previous applications.) In 
Fig.~1, the results of the numerical calculation in that approximation, 
and with the BC vertex (\ref{BC-vertex}), are depicted by the dash-dotted 
line, which practically reaches (from above) the asymptotic behavior, 
although not  as fast as the results without that approximation, depicted 
by the solid line. We now note that in the approximation (\ref{softLimBS}), 
the asymptotic behavior (\ref{ampWtilde}) should be realized with $f_\pi$ 
and ${\widetilde f}_\pi$ calculated in the {\it same} approximation. We 
will denote them by $f_\pi^{gPS}$ and ${\widetilde f}_\pi^{gPS}$, where
${gPS}$ is short for ``generalized Pagels-Stokar". Namely, it turns out 
\cite{jain91} that the soft limit approximation (\ref{softLimBS}), when 
applied to $f_\pi$, is the same as the widely used Pagels-Stokar 
approximation \cite{PagelsStokar79} {\it except} that they work with the 
restriction $A(q^2) \equiv 1$ -- unlike the present approach. Since we 
find ${\widetilde f}_\pi^{gPS}=109.2$ MeV, it appears that our 
dash-dotted curve should approach 
$Q^2T_{\pi^0}^{soft}(-Q^2,0)\to(4/3){\widetilde f}_\pi^{gPS}=0.146$
GeV. In fact, at $Q^2=36$ GeV$^2$,
$Q^2T_{\pi^0}^{soft}(-Q^2,0) \approx  0.136$ GeV.
To understand what forced the dash-dotted curve below, let us 
remember that the factor $1/f_\pi$ appears in Eq. (\ref{softLimBS}) 
because in the chiral limit it turns out that the pion 
decay constant $f_\pi$ is precisely equal to the the normalization of 
the BS vertex $\Gamma_\pi(p,q)$ \cite{JJ}.
In Ref.~\cite{KeKl1} the transition form factor was thus found using
in Eq. (\ref{softLimBS}) our chiral-limit value of the pion decay constant,
$f_\pi = f_\pi^0 = 89.8$ MeV, and to make contact with that reference
we have evaluated in this paper the dash-dotted curve in the same way.
However, from what has just been explained, it follows that when we 
use not only the chiral, but also the soft limit (\ref{softLimBS}),
the normalization of the BS vertex is given by the pion decay constant
calculated in the same approximation, $f_\pi = f_\pi^{gPS}$, which in
our adopted model has the value of 80.8 MeV. The transition form 
factor of Ref.~\cite{KeKl1} was thus suppressed by the factor 
$f_\pi^{gPS}/f_\pi^0$; for the same reason, the dash-dotted line 
in the present Fig.~1 approaches from above 
$(4/3){\widetilde f}_\pi^{gPS}(f_\pi^{gPS}/f_\pi^0)\approx 0.131$ GeV
as the limiting value, so everything tallies.

In the case of the BC vertex, the ``soft" leg adjacent to the pion 
BS-amplitude always contributes $A(q^2)$, which enhances the asymptotic 
value of $Q^2 T_{\pi^0}(-Q^2,0)$. This is not so for the CP or mCP 
vertices. Since they tend to the bare vertex as soon as the high 
momentum flows through one of their legs, $f_\pi$ does {\it not} 
get replaced by ${\widetilde f}_\pi$, and the ``bare" result, 
Eq.~(\ref{largeQ2}) with ${\cal J}=4/3$, continues to hold for 
$Q^2\to\infty$ when the CP or mCP vertices are used. 

Of course, the usage of the mCP vertex
instead of the bare $\gamma^\mu$ makes a considerable difference for 
the finite $Q^2$. Let us therefore now focus our attention to Fig.~3,
which shows $Q^2 T_{\pi^0}(-Q^2,0)$ found in our approach for both the 
BC-case (solid curve) and the mCP-case (dashed curve) for the interval
of the transferred momenta between the presently highest accessible
 $Q^2$ and $Q^2=0$. 

At $Q^2=0$, both BC and mCP vertices give the same amplitude 
(\ref{AnomAmpl}) for $\pi^0\to\gamma\gamma$ in the chiral limit. 
With growing $Q^2$, the BC- and mCP-curves in Fig.~3 must 
nevertheless ultimately separate because of the difference 
$(4/3)({\widetilde f}_\pi - f_\pi)$ of their large-$Q^2$ 
limiting values. At the highest presently accessible momenta,
$Q^2\approx 8$ GeV$^2$, they differ by 9\%, but as we go down 
in $Q^2$, we see that already for $Q^2 \lsim 4$ GeV$^2$, they 
practically coincide. This insensitivity on the choice of the 
WTI-preserving $qq\gamma$-vertex is very indicative. Namely, 
the true solution for the $\Gamma^\mu(p,p^\prime)$ vertex,
the one which we try to imitate by the BC and mCP vertex 
{\it Ans\" atze}, is -- for $Q^2 \lsim 4$ GeV$^2$ -- unlikely to 
give $Q^2 T_{\pi^0}(-Q^2,0)$ significantly different from the 
BC and mCP vertices, since at $Q^2=0$ the transition amplitude 
must (in the chiral limit) again be (\ref{AnomAmpl}),
and at larger $Q^2$ the difference should not be larger than
that resulting from the BC and mCP vertices, as they lead to 
so very different $Q^2 T_{\pi^0}(-Q^2,0)$ as $Q^2\to\infty$.
(One should also keep in mind that the multiplication by $Q^2$
serves only for exposing the asymptotics more clearly, and 
that the difference is even smaller for $T_{\pi^0}(-Q^2,0)$ 
proper, which must in every case decrease as $1/Q^2$.) 
The weak sensitivity (at least for not too large $Q^2$) on 
the WTI-preserving {\it Ansatz} for $\Gamma^\mu(p,p^\prime)$ 
means that high-precision measurements of $T_{\pi^0}(-Q^2,0)$ 
will test unambiguously (and possibly give a hint on how
to improve) the SD and BS model solutions which have so far been 
successful in fitting the low-energy hadron properties such as 
the meson spectrum. For example, our model choice \cite{jain93b} 
obviously (for both BC and mCP vertices) slightly overestimates 
$T_{\pi^0}(-Q^2,0)$ in the region $Q^2 \lsim 4$ GeV$^2$. 

We point out that out of various SDE and BSE solutions in the 
SD-BS approach, once that $f_\pi$ has been correctly reproduced
by $\chi$ [see Eq. (\ref{fpi})], the transition form factor is 
most sensitive on $A(q^2)$, or, more precisely, on its values at 
small and intermediate momenta $-q^2$, where $A(q^2)$ is still 
appreciably different from 1. Eqs. (\ref{ampWtilde})-(\ref{approxFactor})
already explained how $A(q^2)$ drives $Q^2 T_{\pi^0}(-Q^2,0)$ 
upwards through ${\widetilde f}_\pi$ for large $Q^2$ in the case 
of the BC vertex. 

To illustrate what happens at $Q^2 \lsim 8$ GeV$^2$ when the 
$A(q^2)$-profile is decreased, let us enforce by hand the extreme, 
artificial case $A(q^2)\equiv 1$. (To avoid confusion, we stress it 
is for illustrative purposes only, as we cannot have such a SD-solution 
in the adopted approach of Refs.~\cite{jain91,munczek92,jain93b}.)
This leads to the curve traced on Fig.~3 by small crosses,
pertaining to the usage of both BC and mCP vertices (as well 
as the CP ones), since
$A(q^2)\equiv 1$ makes $ \Gamma^\mu_{mCP} \to \Gamma^\mu_{BC}$.
This curve illustrates how the heights of the curves depicting
$Q^2 T_{\pi^0}(-Q^2,0)$ depend on how much the $A(q^2)$-profile
exceeds 1. Obviously, for both the solid curve and the dashed one,
the agreement with experiment would be improved by lowering them
somewhat (at least in the momentum region $Q^2\lsim 4$ GeV),
which could be achieved by modifying the model \cite{jain93b} and/or its
parameters so that such a new solution for $A(q^2)$ is somewhat lowered
towards its asymptotic value $A(q^2\to\infty) \to 1$. (Of course, in
order to be significant, this must not be a specialized re-fitting aimed
{\it only} at $A(q^2)$. Lowering of $A(q^2)$ must be a result of a broad fit
to many meson properties, comparable to the original fit \cite{jain93b}.
This, however, is beyond the scope of this paper.)

By the same token, high precision measurements of $T_{\pi^0}(-Q^2,0)$
(such as those planned at Jefferson Lab \cite{gagasCEBAF}) 
can be especially helpful in
obtaining information on $A(q^2)$ empirically. First note that
Fig.~3 reveals that in the region around $Q^2 \sim 4$
GeV$^2$ we are already in the regime where the dependence on
the BS-amplitude $\chi$ is lumped into $f_\pi$. Therefore,
high-precision measurements of $T_{\pi^0}(-Q^2,0)$ in the
region close to the asymptotic regime will give information
on the integrated strength of $A(q^2)$ although not on $A(q^2)$
itself. However, since it is known that the form of that function
must be a rather smooth transition ({\it e.g.}, see \cite{KeBiKl98})
from $A(q^2)>1$ for $Q^2$ near 0, to $A(q^2)\to 1$ in the $Q^2$-domain 
where QCD is perturbative, such measurements \cite{gagasCEBAF} would 
give a useful hint even about $A(q^2)$ itself -- namely about what
solutions for $A(q^2)$ one may have in sensible descriptions of 
dynamically dressed quarks and their bound states.


\section*{Acknowledgments}
\noindent                 The authors acknowledge the support of the 
           Croatian Ministry of Science Technology to the conference 
        ``Nuclear and Particle Physics with CEBAF at Jefferson Lab", 
                            Dubrovnik, Croatia, November 3-10, 1998.



\section*{Figure captions}

\begin{itemize}

\item[{\bf Fig.~1:}]

Our results for $Q^2 T_{\pi^0}(-Q^2,0)$ are presented from $Q^2=0$
up to $Q^2=36$ GeV$^2$. The data presently exist only for $Q^2<10$
GeV$^2$. CELLO data points are circles and those of CLEO are triangles.
The dash-dotted line represents our $Q^2 T_{\pi^0}(-Q^2,0)$ evaluated 
in the chiral and soft limit approximation (\ref{softLimBS}) and
only with the BC $qq\gamma$ vertices. Our results 
{\it without} that approximation are depicted by the solid line 
for the case of the BC vertex, and by the dashed line for the case of 
the mCP vertex. This latter one gets lower beyond $Q^2 \sim 2.5$ GeV$^2$ 
and does not yet saturate, but at higher $Q^2$ ultimately must approach 
asymptotically $Q^2 T_{\pi^0}(-Q^2,0) \to 4 f_\pi/3$ for higher $Q^2$. 
This limit is also denoted by the dashed line, but the straight one. The 
Brodsky-Lepage interpolation formula \cite{BrodskyLepage} is the dotted 
curve approaching the (also dotted) straight line denoting the pQCD 
limiting value $2 f_\pi$. The line of empty squares denotes basically 
the pQCD prediction (but with fixed $N_f=5$) of Ref.~\cite{Brodsky+al98}.

 \item[{\bf Fig.~2:}]
The pion distribution amplitudes (\ref{genForm}) are depicted in the
dimensionless version $\phi(x)\equiv\varphi_\pi(x)/f_{\pi}$ for the three
values of $\zeta$ pertinent to the discussion in this paper. $\zeta=1$
yields the asymptotic distribution amplitude (the dotted curve),
$\zeta=1.5$ pertains to the pion distribution amplitude
depicted by the solid curve and corresponds to the large-$Q^2$
values of our $T_{\pi^0}(-Q^2,0)$ evaluated with the BC $qq\gamma$
vertex. $\zeta=2.5$ corresponds to the transition form factors
predicted by Refs.~\cite{Brodsky+al98} and \cite{Radyushkin+Rusk3}
for the largest $Q^2$-values ($Q^2\approx 8$ GeV$^2$) at which 
$T_{\pi^0}(-Q^2,0)$ has been measured so far.

\item[{\bf Fig.~3:}]
The comparison of our results for the pion transition form factor
(times $Q^2$) with the CELLO (circles) and CLEO (triangles) data.
Our results for $Q^2 T_{\pi^0}(-Q^2,0)$ 
are depicted by the solid line for the case of the BC vertex,
and by the dashed line for the case of the mCP vertex. 
This latter one gets lower
beyond $Q^2 \sim 2.5$ GeV$^2$ and does not yet saturate at the
presently accessible momenta although approaches asymptotically
$Q^2 T_{\pi^0}(-Q^2,0) \to 4 f_\pi/3$ for much higher $Q^2$.
(This limit is denoted by the dashed straight line.)
The little crosses denote (for both BC and mCP $qq\gamma$ vertices)
our $Q^2 T_{\pi^0}(-Q^2,0)$ when we enforce $A(q^2)\equiv 1$ by hand.

\end{itemize}


\newpage

\vspace*{4cm}
\epsfxsize = 16 cm \epsfbox{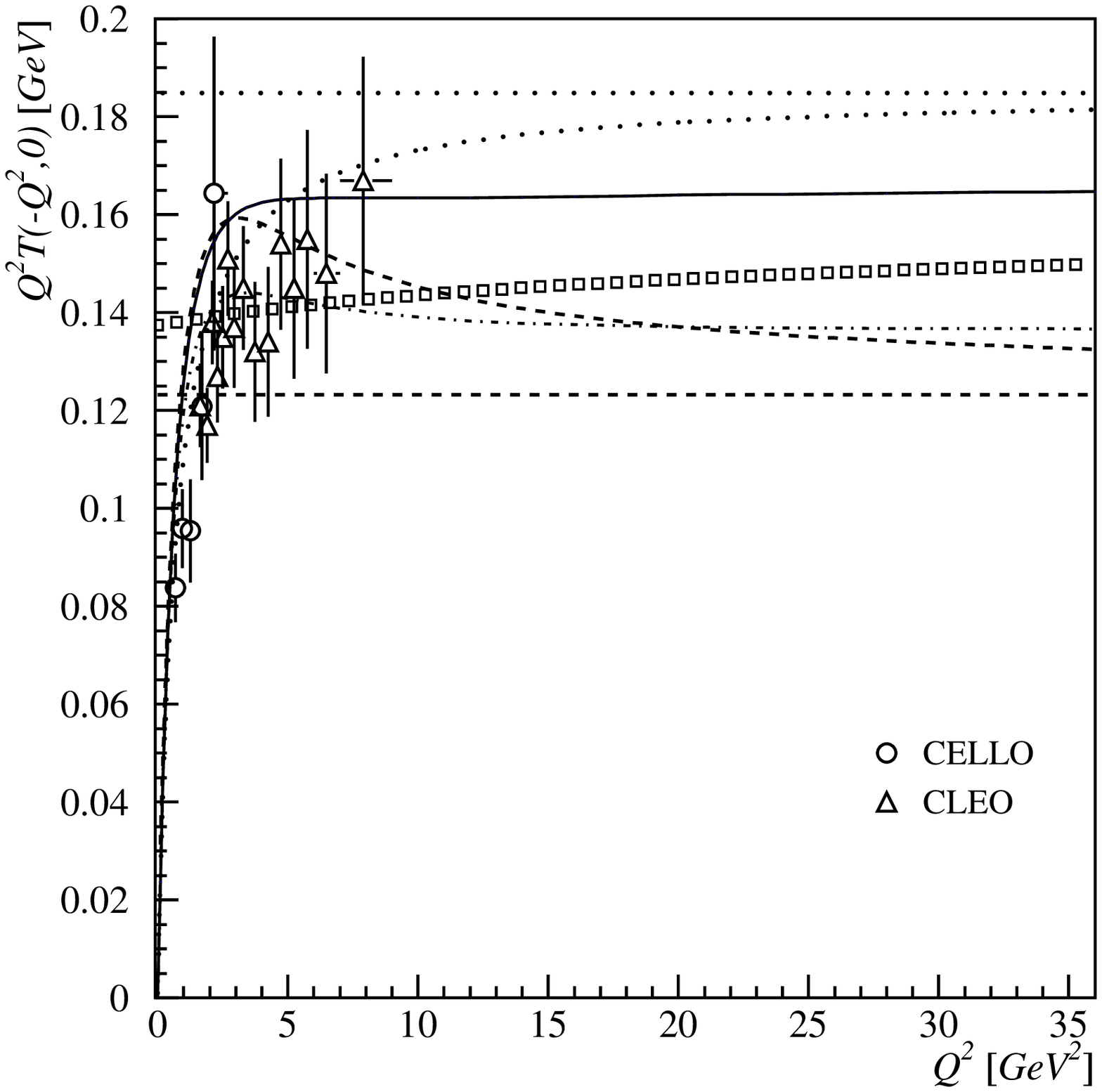}

\newpage

\vspace*{2cm}
\epsfxsize = 16 cm \epsfbox{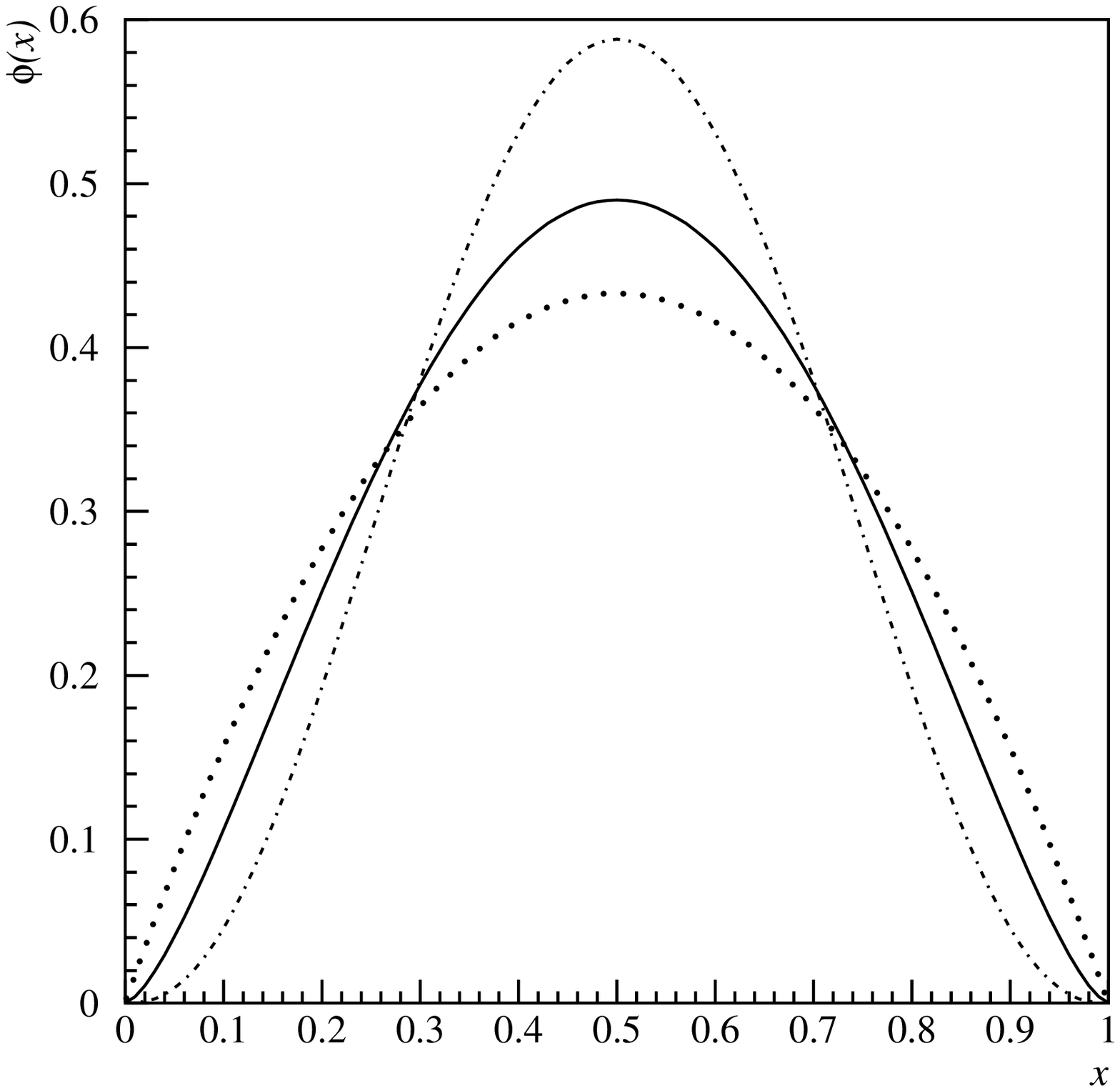}

\newpage

\vspace*{2cm}
\epsfxsize = 16 cm \epsfbox{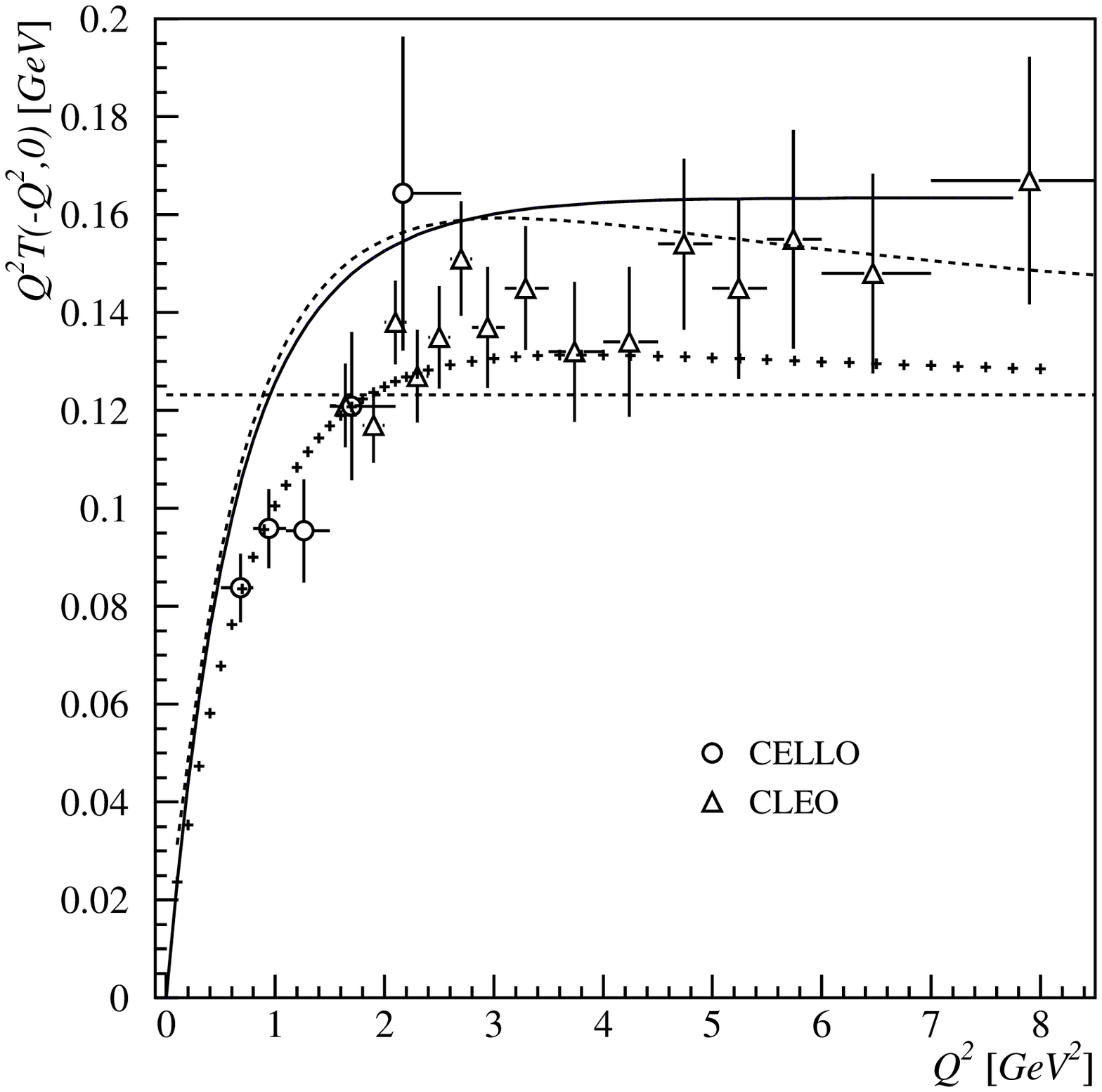}

\end{document}